\RequirePackage[2020-02-02]{latexrelease}
\documentclass[%
 aip,
 amsmath,amssymb,
reprint,%
]{revtex4-1}

\usepackage{graphicx}
\usepackage{dcolumn}
\usepackage{bm}
\usepackage{xcolor,soul}
\usepackage{comment}

\usepackage[mathlines]{lineno}

\begin{document}

\title{Stationary states in two lane traffic: insights from kinetic theory}
\author{A. Sai Venkata Ramana}
\affiliation{New York University Abu Dhabi, Saadiyat Island, P.O. Box 129188, Abu Dhabi, U.A.E.}
\author{Saif Eddin Jabari}
\affiliation{New York University Abu Dhabi, Saadiyat Island, P.O. Box 129188, Abu Dhabi, U.A.E.}
\affiliation{New York University Tandon School of Engineering, Brooklyn NY.}
\date{\today}
\begin{abstract}
Kinetics of dilute heterogeneous traffic
on a two lane road is formulated in the framework of Ben-Naim Krapivsky model and stationary state properties
are analytically derived in the asymptotic limit.
The heterogeneity is introduced into the model as a quenched disorder in
desired speeds 
of vehicles. The two-lane model assumes that each
vehicle/platoon in a lane moves ballistically
until it approaches a slow moving vehicle/platoon and then
joins it. Vehicles in a platoon are assumed to escape 
the platoon at a constant rate by 
changing lanes after which they continue to 
move at their desired speeds. 
Each lane is assumed to have a different escape rate.
As the stationary state is approached, the platoon
density in the two lanes become equal,
whereas the vehicle densities and fluxes are higher in the lane
with lower escape rate. 
A majority of
the vehicles enjoy a free-flow if the harmonic mean of the escape
rates of the lanes is comparable to 
average initial flux on the road. The average platoon size is close 
to unity in the free-flow regime. If the harmonic mean is lower than the average initial flux, then vehicles
with desired speeds lower than a characteristic speed $v^*$ still
 enjoy free-flow while those vehicles with desired speeds that are greater
 than $v^*$ experience congestion and form platoons behind
 the slower vehicles. The characteristic speed depends on
 the mean of escape times $(R=(R_1+R_{-1})/2)$ of the two lanes (represented by 1 and -1) as
   $v^* \sim R^{-\frac{1}{\mu+2}}$, where $\mu$ is the exponent 
   of the quenched disorder distribution for desired speed 
 in the small speed limit. The average platoon size in a lane, when $v^* \ll 1$, is 
 proportional to $R^{\frac{\mu+1}{\mu+2}}$ plus a lane dependent 
 correction. Equations for the kinetics of platoon size distribution
 for two-lane traffic are also studied. 
  It is shown that a stationary state
 with platoons as large as road length can occur 
 only if the mean escape rate is independent of platoon
 size.

\end{abstract}

\maketitle

\section{Introduction}
Research on the development 
  of algorithms related to mobility of connected and automated vehicles (CAVs) has taken a 
  front seat as it is expected that a traffic system with CAVs would help in efficient 
  traffic management, reduce energy consumption and pollution,  apart from other advantages like increasing
  safety etc\cite{Habib, Stevens}.
An important application of these algorithms for CAVs is cooperative driving automation (CDA)\cite{SAE_J3216, Opencda}
by which vehicles operate and move in a cooperative way by sharing information.
 CDA brings in a need to understand the Physics of heterogeneous traffic in a detailed way, especially 
the collective phenomena that occur in heterogeneous traffic, for building efficient algorithms to implement the CDA. 
 
Models for traffic on a single lane predict the occurrence of 
interesting collective phenomena in the kinetics of relaxation to
stationary state and in some physically observable quantities in the
stationary state when passing is not allowed\cite{ Krug1996, Evans1996, Ktitarev1997,Bengrine1999, DC2000, Barma2006, Ramana2020, Ramana2021, Ben1994}.
The main reason for the
emergence of the collective phenomena is heterogeneity in traffic (namely, differences in driving characteristics from vehicle to vehicle), typically modeled as quenched disorders in the parameters of the
models\cite{Krug1996,Ktitarev1997,Saif2014,Saif2018}. For instance, an interesting phenomenon occurs
when the desired speeds of the drivers/vehicles in the traffic
are heterogeneous and if passing is not allowed. Faster vehicles in the system form platoons behind the slower vehicles
and the average platoon size grows with time
as a power-law  $ \sim t^{\mu+1/\mu+2}$ until all the platoons coalesce 
and form a single giant platoon. $\mu$ is the exponent of the quenched disorder 
distribution in desired speeds of vehicles in the low-speed limit.
The collective phenomena on a single lane are have been studied
using various kinds of approaches of cellular-automata,
car-following models\mbox{\cite{DC2000,Helbing2001,Han2021}}. New approaches are still being developed
 to efficiently model the complex phenomena. A particularly
  interesting development is the application of evolutionary games
  to traffic\mbox{\cite{Tanimoto1,Tanimoto2}}.
Nevertheless, the question of what happens in a two lane system,
when lane changing is allowed, is not fully understood.

 Ben Naim, Krapivsky and Redner developed a kinetic theory
 for which they could analytically derive the power-law for growth of platoon size with time\cite{Ben1994}. 
In a series of later works \cite{Ben1997, Ben1998, Ben1999, Ispolatov2000},
Ben-Naim and Krapivsky (BK) included passing of vehicles in their formalism
and came up with an equation similar to the Boltzmann equation.
They studied the stationary states of the system and showed that 
the platoons of vehicles 
still form when the rate of escape from a platoon is smaller than the
rate of formation of the platoon. 
The beauty of the formulation by BK is that, despite its simplifying assumptions, some analytical
results could be obtained which give a qualitative understanding of the phenomenon.
However, the formalism
 was developed for a single lane. A more realistic case would be 
 to consider two lanes with vehicles allowed to change lanes at any 
 point on the road.
 The nuances of a two-lane picture
 over a single-lane picture needs to be understood as that would provide 
 insights into collective phenomena in road traffic in a more realistic 
 way. Thus, in the present work we formulate the two-lane road problem 
 within the framework laid out by the BK model and study its stationary 
 states. 
 
 In Sec.~\ref{kin}, the kinetic equations for velocity distribution of platoons
 and vehicles for a two lane traffic are formulated. 
 General forms for the velocity distributions in the stationary state are 
 derived in Sec.~\ref{stst}. Conditions for a majority of vehicles 
 to experience free flow and those for congested flow
 are derived in Sec.~\ref{ffc}
  and various useful quantities are obtained in the asymptotic limits.
 In Sec.~\ref{ppt},
 the equations for platoon size distribution in a two lane traffic system
 are formulated 
 and the platoon phase transition is studied. The results are summarized 
 in Sec.~\ref{sd}.

\section{Kinetic theory for two lane traffic flow } \label{kin}
 Consider an infinitely long two-lane freeway with lane changing allowed in any 
part of the road.  Suppose the traffic is heterogeneous due to 
a quenched disorder in the (free-flow) desired speeds of vehicles. 
All vehicles, at $t=0$, move with their desired 
speeds which are drawn independently from a bounded probability 
density $P_{0}(V_d=v)$. 
Assume that the length scale considered is 
sufficiently large that the probability distribution is accurately sampled by the 
vehicles in one unit of length. 
As the traffic flow evolves in time,
formation and dissolution of
platoons (clusters) of vehicles happens. A platoon's size grows when a fast moving
platoon joins a slow moving platoon and the two platoons merge; we will refer to this merging of platoons as a ``collision'' event (not to be confused with an actual collision between two vehicles). 
The platoons move ballistically between successive collisions.
A platoon's size drops when a vehicle in the platoon 
changes lanes, which we refer to as an ``escape'' event. 
After changing lanes the vehicle moves at its desired
speed until it collides with another vehicle in that lane. The 
vehicles are assumed to instantaneously increase or decrease their
speeds which is justified for a dilute system 
where the mean collision time is much larger than the mean relaxation time of speed.

 We use the same dimensionless units as used by BK for consistency. 
  The two lanes are indexed as $-1$ for one of the lanes and $1$ for the other; theses indices appear as suffixes throughout the paper. 
  In the dimensionless 
  units, velocity is scaled by $v_0$ i.e., $v/v_0 \mapsto v$ where $v_0$ is the 
  maximum free-flow speed available, i.e., $v_0 \equiv \sup_v P_{0}(v)$.
  Let $\rho_0 = (\rho_{01} + \rho_{02})$ be the initial \emph{global} density 
   of vehicles on the road (total number of vehicles per unit length of the system), where $\rho_{0i}$ is the initial global density in lane $i$.  $\rho_0$ remains constant over time; it's inverse, $\overline{s} = \rho_0^{-1}$, is the initial mean gap between vehicles on the road.
   We use $\overline{s}$ to scale the spatial coordinates as
   $x \overline{s}^{-1} = x \rho_0 \mapsto x$ to make it dimensionless.
   Similarly time is scaled as $\rho_0 v_0 t \mapsto t$ which renders $t$ dimensionless. $(\rho_0 v_0)^{-1}$ may be interpreted as initial mean platoon collision time.
   Let $R_i$ be the mean escape time of vehicles in lane $i$. In dimensionless units, $\rho_0 v_0 R_i \mapsto R_i$.
   All the speeds are relative to the
   slowest vehicle in the system.
   
   
   For book keeping purposes,
   we term those vehicles moving at their desired speed $v$ as $v$-vehicles and those vehicles whose desired
   speed is $v$ but are moving at a slower speed $v'$ are termed
   as $(v,v')$-vehicles. 

Let $\Phi_{i}(v,t)$ ($i \in \{-1,1\}$) be  
 the density of platoons (of any size) moving at 
 speed $v$ which are led by vehicles whose desired speed is $v$
 (i.e., $v$-vehicles).   Thus, it also gives the density of \emph{vehicles} moving at their desired speed $v$.
 At time $t=0$, we have that $\Phi_i(v,0) = P_{0}(v) \rho_{0i}$, 
 implying that initially there are only single vehicles (platoons of 
 size $1$) and all start moving at their desired speed. 
Let $\Psi_i(v',t;v)$ be the  
density of $(v,v')$-vehicles at time $t$.
 The speeds of these vehicles would have dropped as a result of 
 collisions with slower platoons.
 The master 
 equation describing the evolution of platoon density in lane $i$ is
 \begin{multline}
     \frac{\partial \Phi_i(v,t)}{\partial t} = \frac{1}{R_{-i}} \int_0^v dv' \Psi_{-i}(v',t;v) \\- \Phi_i(v,t)\int_0^v dv'(v-v')\Phi_i(v',t). \label{Cd1}
 \end{multline}
The first term on the right-hand side (RHS) represents the rate at 
which $\Phi_i(v,t)$ increases as a result of 
 $(v,v')$-vehicles in the adjacent lane ($-i$) changing lanes to 
 increase their speed to
 their desired speed $v$. $R_{-i}^{-1}$ is the average rate at which 
 vehicles leave lane $-i$. $\int_0^v dv' \Psi_{-i}(v',t;v)$ is the 
 density of vehicles in lane $-i$ moving slower than their desired 
 speed $v$. 
 per unit time from lane $-i$ to lane $i$. 
 The second term represents the 
 number of collisions per unit length 
 per unit time in lane $i$ between platoons moving at speed $v$ and slower 
 platoons ahead, thereby slowing down.  The second term employs 
 Boltzmann's \emph{Stosszahlansatz} assumption (a.k.a. molecular 
 chaos). It essentially states that the rate of collision is 
 proportional to the difference in speed, the higher the difference, 
 the more likely the collision (or the higher the rate of collision). 
 The master equation describing the evolution of slowed-down cars in 
 lane $i$ is 
 \begin{multline}
 \frac{\partial \Psi_i(v',t;v)}{\partial t} = -\frac{1}{R_i}\Psi_i(v',t;v) \\
 - \Psi_i(v',t;v)\int_0^{v'} dw(v'-w)\Phi_i(w,t) \\
 + \Phi_i(v',t)\int_{v'}^v dw (w-v')\Psi_i(w,t;v) \\
  + \Phi_i(v,t)\Phi_{i}(v',t)(v - v'). 
  \label{P2v}
 \end{multline}
The first term on the RHS represents the 
rate at which $(v,v')$-vehicles escape a platoon
by changing from lane $i$ to lane $-i$. 
The second term represents collisions between $(v,v')$-vehicles
and platoons moving at speeds lower than $v'$.
 The third term represents collisions between platoons moving at a speed of $v'$ and
  $(v,w)$-vehicles for which
  $v' < w < v$. 
  The fourth term is the collision rate between $v$-platoons and $v'$-platoons. Note that there is no term due to vehicles from
  lane $-i$ as it is assumed that the vehicles attain their
  desired speed as soon as they change lane.
  Equations \eqref{Cd1} and \eqref{P2v} are a set of four coupled equations that describe the evolution of a two lane traffic within the BKR model.
  The initial conditions for Equations \mbox{\eqref{Cd1} and \eqref{P2v}} are $\Phi_i(v,0) = P_{0}(v)\rho_{0i}$ and 
  $\Psi_i(v',0;v)=0$, respectively. 
  The density of vehicles in lane $i$ whose desired speed is $v$ is
  \begin{equation}
      \mathfrak{N}_i(v,t) = \Phi_i(v,t) + \int_0^v dv' \Psi_i(v',t;v).
      \label{Pf}
  \end{equation}
  The density of vehicles on any lane whose desired speed is
  $v$ is
  $\mathfrak{N}(v,t) = \frac{1}{2}(\mathfrak{N}_{-1}(v,t) + \mathfrak{N}_1(v,t))$. From Equations \mbox{\eqref{Cd1} and \eqref{P2v}} we have that
  \begin{equation}
      \frac{\partial{\mathfrak{N}(v,t)}}{\partial t} = 0, \label{cons}
  \end{equation}
  which implies that
  \begin{equation}
     \mathfrak{N}_{-1}(v,t)+ \mathfrak{N}_1(v,t) = P_0(v) \label{Pfc}
  \end{equation}
  for all $t \ge 0$, confirming the conservation of the number of vehicles by the equations of the model. To show that \mbox{\eqref{cons}} is true, we only need to note that the second and third terms in \mbox{Eq.\eqref{P2v}} cancel one another upon integrating $v'$ over $(0,v)$, that is,
  \begin{multline*}
    - \int_0^v \int_0^{v'} dv' dw (v'-w) \Psi_i(v',t;v) \Phi_i(w,t) \\
 + \int_0^v  \int_{v'}^v dv' dw (w-v')\Psi_i(w,t;v) \Phi_i(v',t) = 0,
  \end{multline*}
  since the domain of integration of the first integral, $\{(v',w): v'\in(0,v), w \in (0,v')\}$, is a triangular region in $\mathbb{R}^2$ that can be equivalently written as $\{(w,v'): w \in (0,v), v' \in (w,v) \}$.  For the first integral above, this equivalence implies that
  \begin{multline*}
    \int_0^v \int_0^{v'} dv' dw (v'-w) \Psi_i(v',t;v) \Phi_i(w,t) \\= \int_0^v \int_w^v dw dv' (v'-w) \Psi_i(v',t;v) \Phi_i(w,t).
  \end{multline*}
  A change of variable shows that the terms cancel one another as claimed. The remaining steps needed to demonstrate that \mbox{E.\eqref{cons}} follow immediately from the definitions.
  
   In the next section we
   study the stationary state behavior of the system of equations Eq.\eqref{Cd1}, Eq.\eqref{P2v} and the Eq.\eqref{Pfc}. 
  
\section{The stationary state} \label{stst}
As the stationary state is approached,  ``gain" becomes
equal to ``loss" and all the densities become time independent.
Thus, in each of the Eq.\eqref{Cd1} and 
Eq.\eqref{P2v}, LHS tends to zero asymptotically in time.
Therefore, we have in the stationary state 
\begin{equation}
   \frac{1}{R_{-i}} \int_0^v dv' \Psi_{-i}(v';v) = \Phi_i(v)\int_0^v dv'(v-v')\Phi_i(v') \label{Esc1eqcol1}
 \end{equation}
 and
\begin{multline}
\frac{1}{R_i}\Psi_i(v';v) = - \Psi_i(v';v)\int_0^{v'} dw(v'-w)\Phi_i(w) \\
 +  \Phi_i(v')\int_{v'}^v dw (w-v')\Psi_i(w;v) 
  \\ + \Phi_i(v)\Phi_{i}(v')(v - v'). \label{P2vs}
 \end{multline}
 Note that we have dropped the dependence on $t$ in this stationary setting. Integrating Eq.\eqref{P2vs} on both sides over $v'$ between $0$ and $v$,
we get 
\begin{equation}
   \frac{1}{R_i}\int_0^v dv' \Psi_i(v';v) = \Phi_i(v)\int_0^v dv'(v-v')\Phi_i(v'), \label{Esc1eqcol2}
\end{equation}
where we have again used the fact that the first and second terms in \mbox{Eq.\eqref{P2vs}} cancel each other upon integrating $v'$ over $(0,v)$.  For convenience, we define the auxiliary functions
\begin{equation}
    Q_i(v) = R_i^{-1} + \int_0^v dv' (v - v')\Phi_i(v') \label{Qi}
\end{equation}
and
\begin{equation}
    {\widetilde{Q}_i(v';v)} = \int_{v'}^v dw (w-v') \Psi_i(w;v). \label{Q2i}
\end{equation}
Then using Eq.\eqref{Esc1eqcol2} and Eq.\eqref{Pf}, we can write Eq.\eqref{Pfc} as
 \begin{equation}
     R_1\Phi_1(v)Q_1(v) + R_{-1}\Phi_{-1}(v)Q_{-1}(v) = P_0(v). \label{Probcons}
 \end{equation}

Equating Eq.\eqref{Esc1eqcol2} and Eq.\eqref{Esc1eqcol1},
we get
\begin{equation}
  \Phi_1(v) \xi_1(v) = \Phi_{-1}(v)\xi_{-1}(v)  \label{Col1eqcol2}
\end{equation}
where $\xi_i(v) = Q_i(v)-R_i^{-1}$ is the stationary collision rate.
Eq.\eqref{Esc1eqcol2} and Eq.\eqref{Esc1eqcol1} mean that  the escape rate and the collision rate on any lane must become equal
 in the stationary state. 
Noting that 
\begin{equation}
    Q_i''(v) = \frac{\partial^2 Q_i(v)}{\partial v^2} = \Phi_i(v)
\end{equation}
and
\begin{equation}
   \widetilde{Q}''_i(v';v) = \frac{\partial^2 \widetilde{Q}_i(v';v)}{\partial v'^2} = \Psi_i(v';v)
\end{equation}
with $Q_i(0)=R_i^{-1}$, $Q_i'(0) =0$, $\widetilde{Q}_i(v;v)=0$ and 
$\frac{\partial \widetilde{Q}_i(v; v)}{\partial v'}=0$, Eq.\eqref{Probcons} and \mbox{Eq.\eqref{Col1eqcol2}} 
may be re-written as
 \begin{equation}
     R_1Q_1''(v)Q_1(v) + R_{-1}Q_{-1}''(v)Q_{-1}(v) = P_0(v) \label{Probcons2}
 \end{equation}
 and
\begin{equation}
  Q_1''(v)\xi_1(v) = Q_{-1}''(v)\xi_{-1}(v)  \label{Col1eqcol22}
\end{equation}
Eq.\eqref{Probcons2} and
Eq.\eqref{Col1eqcol22} describe the density of platoons moving with speed $v$
 in the stationary state.
 The density of vehicles moving with speed $v$ is given by 
 \begin{equation}
     G_i(v) = \Phi_i(v) + \int_v^1 dw \Psi_i(v;w). \label{carden}
 \end{equation}
 Using Eq.\eqref{Qi} and Eq.\eqref{Q2i} in Eq.\eqref{P2vs} we get
 \begin{multline}
     \frac{\partial^2 \widetilde{Q}_i(v';v)}{\partial v'^2}Q_i(v')= \Phi_i(v)\Phi_i(v')(v-v') \\
     + \frac{\partial^2 Q_i(v')}{\partial v'^2}\widetilde{Q}_i(v';v) \label{P2vs2}
 \end{multline}
 Integrating the above equation using the boundary conditions gives
 \begin{equation}
     \Psi_i(v';v) = \Phi_i(v)Q_i(v)\Phi_i(v')\int_{v'}^v \frac{du}{[Q_i(u)]^2} \label{P2vs3}.
 \end{equation}
 Equations \eqref{Probcons2}, \eqref{Col1eqcol22} and \eqref{P2vs3}
 completely describe the stationary state of the system. Upon solving them,
  $\Phi_i(v)$, $\Psi_i(v';v)$ can be obtained, and from these other
  quantities of interest may be obtained. A recipe for calculating a few quantities 
  is as follows: the density of platoons in 
  lane $i$ is 
  \begin{equation}
      \rho_i = \int_0^1dv \Phi_i(v)
  \end{equation}
  and the total platoon density is
  \begin{equation}
      \rho = (\rho_1 + \rho_{-1})/2.
  \end{equation}
  The average platoon size $\Lambda_i$ is 
  \begin{equation}
      \Lambda_i \sim \rho_i/\rho_i^c
  \end{equation}
  where 
  \begin{equation}
      \rho_i^c = \int_0^1dv~G_i(v).
  \end{equation}
  The average platoon speed is given by
  \begin{equation}
      \langle v_i \rangle = \rho_i^{-1}\int_0^1dv~v ~\Phi_i(v).
  \end{equation}
  The flux in  lane $i$ is given by
\begin{equation}
    J_i = \int_0^1 dv~v~G_i(v).
\end{equation}
\section{Free flow and congestion} \label{ffc}
As seen in the single lane case without passing, even in 
dilute traffic, vehicles with higher desired speeds
face congestion and those with lower desired speed always 
experiences free-flow.  When passing is allowed in the single lane, 
Ben Naim and Krapivsky showed that vehicles having their desired speed 
below a characteristic speed $v^*$ experience free-flow while those 
having desired speed above $v^*$ face congestion.
Below we quantify the phenomenon for the two lane case with lane changing. 

First we note that $\Phi_i$ not only represents
the density of 
platoons in lane $i$ moving with speed $v$ but it also represents
the density of $v$-vehicles in lane $i$.
A majority of these vehicles 
experience free-flow in the stationary state if, in Eq.\eqref{Probcons},
\begin{equation}
   0 \sim R_1\Phi_1(v)\xi_1(v) + R_{-1}\Phi_{-1}(v)\xi_{-1}(v)  \ll \Phi_1(v) + \Phi_{-1}(v),
   \label{cclineq}
\end{equation}
 which basically means that the total number of collision events
experienced by $v$-vehicles on both the lanes 
over a time scale of their respective escape times
is much less than the number of $v$-vehicles.
Above condition occurs when $R_1\xi_1(v)$ and $R_{-1} \xi_{-1}(v)$ are small.
As $\xi_i(v)$ is a monotonically increasing function of $v$, the smallness
 of $\xi_i(v)$ implies $v$ has to be small. Thus the requirement that
 $R_i\xi_i(v)$ has to be small implies that $v$ has to be smaller than a 
 characteristic speed $v^*$ as shown below. When above condition is satisfied, 
Eq.\eqref{Probcons} becomes
\begin{equation}
   \Phi_1(v) + \Phi_{-1}(v) \approx P_0(v). \label{Probconsccl}
\end{equation}
Using Eq.\eqref{Col1eqcol2} in above equation, we get
\begin{equation}
    \Phi_i(v) \approx \frac{P_0(v)\xi_{-i}(v)}{\xi_1(v) + \xi_{-1}(v)}
\end{equation}
A solution which satisfies both Eq.\eqref{Probconsccl} and Eq.\eqref{Col1eqcol2} is
$\xi_1(v) = \xi_{-1}(v)$ which implies
\begin{equation}
    \Phi_1(v) = \Phi_{-1}(v) \approx P_0(v)/2. \label{sol0ccl}
\end{equation}
Considering the above forms of $\Phi_1$ and $\Phi_{-1}$ as zeroth order terms 
of a perturbation series with $(R_1 + R_{-1})\xi_0(v)$ as a perturbation parameter,
one may obtain higher order terms in the series 
 for $\Phi_1$ and $\Phi_{-1}$,  
 where $\xi_0(v) = \int_0^v dv'(v - v') P_0(v')$. The choice of the 
 perturbation parameter may be motivated by Eq.\eqref{cclineq}, which says
 $(R_1 + R_{-1})\xi_0(v) \sim 0$. 
 Using the zeroth order forms of $\Phi_i$ in $\xi_i$, in \mbox{Eq.\eqref{Probcons}} 
 and Eq.\eqref{Col1eqcol2}, the first order approximation turns out to be
\begin{equation}
    \Phi_i(v) \approx \frac{P_0(v)}{2}\left( 1 - \frac{R_1 + R_{-1}}{4}\xi_0(v) \right).
    \label{Pi1o}
\end{equation}
When $v \sim v^*$,
 \begin{eqnarray}
 R_1\Phi_1(v)\xi_1(v) + R_{-1}\Phi_{-1}(v)\xi_{-1}(v) \sim \Phi_1(v) + \Phi_{-1}(v)
 \end{eqnarray}
 Approximating $\Phi_1(v)\sim \Phi_2(v) \sim P_0(v)/2$, we get
 \begin{equation}
  \int_0^{v^*} dv (v^*-v)P_0(v) \sim  \left(\frac{R_1 + R_{-1}}{4}\right)^{-1}.
 \end{equation}
If  $P_0(v) \sim Av^{\mu}$ for $v \in [0,v^*)$ we see
 that
 \begin{equation}
     v^* \sim \left(\frac{R_1 + R_{-1}}{4}\right)^{-1/(\mu + 2)}, \label{v*}
 \end{equation}
 where $\mu>-1$ to ensure normalizability of $P_0(v)$.We may also arrive at the above form of $v^*$ 
 by equating the perturbation
 parameter to unity. For those $v$-vehicles with $v > v^*$, 
 the perturbation parameter is no longer small and the perturbation
 series becomes invalid. Both lanes have same $v^*$ as expected 
 because of mixing of vehicles. If 
   $v^* < 1$, then most of vehicles with 
  their desired speed between $v^*$ and $1$ (maximum speed) would have
   have slowed down due to collisions and
   experience congestion as the stationary state is reached.
   The power-law decay of $v^*$ with $R_1 + R_{-1}$ would be 
   faster for smaller values of $\mu$.
   $v^* > 1$ actually signifies that all the vehicles on the road would experience
   free-flow irrespective of their speed as the maximum available speed is $1$.
   The conditions $v^*\sim 1$ separating the free-flow regime from the mixed regime (with both congested and free-flowing traffic) 
   may be understood in a more illuminating way by writing $v^*$ in dimensional form,
    which is 
   \begin{multline}
              v^* \sim (\rho_0 v_0)^{-1/(\mu + 2)} \left(\frac{R_1 + R_{-1}}{4}\right)^{-1/(\mu + 2)} \\= \left( \frac{\rho_0 v_0}{2} \left(\frac{R_1 + R_{-1}}{2} \right) \right)^{-1/(\mu + 2)}
    \end{multline}
   Therefore,
   \begin{equation}
        v^* \sim 1 
          \implies  \frac{2}{R_1 + R_{-1}} \sim \frac{\rho_0 v_0}{2},
           \label{vsdim}
   \end{equation}
   where $\rho_0 v_0/2$ is the lane averaged initial flux and $2/(R_1 + R_{-1})$ is harmonic
   mean of escape rates ($R_1^{-1}$ and $R_{-1}^{-1}$) of individual lanes. Hence, 
   a majority of the vehicles in the system enjoy free-flow if the harmonic mean of the escape rates on the lanes is of the order of average initial flux. Therefore,
   if the initial density is very low and/or the maximum desired speed of the
   drivers is very small, then the system experiences a free-flow even if the escape
   rates on the lanes are low. On the other hand, if the initial density is high
   and the maximum desired speed is also high, the escape rates have to be high
   to maintain free-flow. 
   Below we calculate some quantities of
   interest for both the cases $v^*>1$ and $v^* < 1$.

\subsection{Case: $v^* > 1$} \label{vs>1}
The total platoon density $\rho_i$ in lane $i$ to first order in $(R_1 + R_{-1})$ is
\begin{equation}
    \rho_i \approx \frac{1}{2} - \frac{R_1 + R_{-1}}{4}\int_0^1 dvP_0(v)\xi_0(v). \label{ccon}
\end{equation}
Thus, mixing of vehicles between
lanes due to lane changing equalizes the platoon density on both the lanes
 as the stationary state is reached.
 
Using $\Phi_i(v) \approx P_0(v)/2$ in $Q_i$ and noting
that $(R_1 + R_{-1})\xi_0(v) \ll 1 $, we find that $Q_i(v) \approx R_i^{-1}$
which on substitution into Eq.\eqref{P2vs3} gives
\begin{equation}
    \Psi_i(v';v)\approx \frac{R_i}{4}P_0(v)P_0(v')(v - v') \label{P2vsccl}
\end{equation}
 Using Eqs.\eqref{Pi1o} and \eqref{P2vsccl} in Eq.\eqref{carden} gives 
 \begin{multline}
G_i(v) \approx \frac{P_0(v)}{2} -\frac{R_1+R_{-1}}{8}P_0(v)\xi_0(v) \\+ \frac{R_i}{4}P_0(v)\int_v^1 dv'(v'-v)P_0(v').
 \end{multline}
 The total vehicle density in lane $i$ is
 \begin{multline}
     \rho_i^c = \int_0^1 dv G_i(v) \approx \frac{1}{2} + \frac{R_i-R_{-i}}{8}\int_0^1 dv P_0(v)\xi_0(v) \\
              = \rho_{0_i}+ \frac{\rho_{0_{-i}}-\rho_{0_i}}{2} + \frac{R_i-R_{-i}}{8}\int_0^1 dv P_0(v)\xi_0(v). \label{rhoic}
 \end{multline}
 The average platoon size in lane $i$ is
 \begin{equation}
 \Lambda_i = \rho_i^c/\rho_i \approx 1 + \frac{3 R_i + R_{-i}}{4} \int_0^1 dv P_0(v)\xi_0(v),
 \end{equation}
 which is close to unity as $R_1$ and $R_2$ are small.
 The $G_i(v)$ obtained above can be used to 
calculate the flux on lane $i$ which turns out to be 
\begin{multline}
    J_i \approx \frac{J_0}{2} + \frac{R_i-R_{-i}}{8}A_0 \\ 
      = J_{0_i} + \frac{\rho_{0_{-i}}-\rho_{0_i}}{2} J_0 + \frac{R_i-R_{-i}}{8}A_0 ,  \label{Fluxccl} 
\end{multline}
where $J_0$ is the first moment of $P_0(v)$ i.e.,
\begin{equation}
    J_0 = \int_0^1 dv v P_0(v), \nonumber
\end{equation}
\begin{equation}
    A_0 = \int_0^1 dv v P_0(v)\xi_0(v), \nonumber
\end{equation}
and 
\begin{equation}
    J_{0_i} = \rho_{0_i}\int_0^1 dv v P_0(v) \nonumber
\end{equation}
is the initial flux. 
Thus,
the total vehicle density and the flux on each lane are different.
For instance, if $\rho_{0_1}>\rho_{0_{-1}}$ and $R_{-1} < R_1 < 1$, then
Eq.\eqref{rhoic} and Eq.\eqref{Fluxccl} say that $\rho_1^c > \rho_{-1}^c$ and $J_1 > J_{-1}$
which is
physically expected as vehicles leave lane $1$ at a slower rate than those entering
lane $1$.
 
To summarize, when $v^* > 1$, the system reaches to a stationary state
in which both the lanes have same platoon density which is similar to that
derived for a single lane with an effective initial velocity distribution
$P_0(v)/2$ and 
effective escape time $(R_1 + R_{-1})/2$ i.e., the mean of the escape times
of both the lanes.
However, the vehicle density and flux on each road is different from
that of the single lane case and thus the two lane case cannot be exactly 
mapped to the single lane case. 
\subsection{Case: $v^*<1$} \label{vs<1}
Those $v$-vehicles with $v<v^*$ experience free-flow and the
above discussed perturbation series is valid for $\Phi_i(v)$.
For $v$-vehicles with $v \gg v^*$,
\begin{equation}
   R_1\Phi_1(v)\xi_1(v) + R_{-1}\Phi_{-1}(v)\xi_{-1}(v)  \gg \Phi_1(v) + \Phi_{-1}(v). \label{eclineq }
\end{equation}
As $v^* \ll 1$ implies $(R_1 + R_{-1}) \gg 1$ from Eq.\eqref{v*}.
Using the above approximation and Eq.\eqref{Col1eqcol2} in Eq.\eqref{Probcons},
$\Phi_1(v)$ and $\Phi_{-1}(v)$ get decoupled and satisfy
\begin{equation}
    (R_1 + R_{-1})\Phi_i(v)\xi_i(v) \approx P_0(v)
\end{equation}
or
\begin{equation}
    \xi_i''(v)\xi_i(v) \approx P_0(v)(R_1 + R_{-1})^{-1} \approx P_0(v) v^{*^{\mu+2}}. \label{Probconsecl}
\end{equation}
 In the above form, the equations for platoon density are same as those 
 of single lane case with an effective escape rate $(R_1 + R_{-1})^{-1}$. Looking at
 the equations, it may be inferred that $\Phi_1 = \Phi_{-1}$ in this case as well. Ben Naim
  and Krapivsky derived an approximate form for $\Phi_i$ using Eq.\eqref{Probconsecl} when 
  \begin{equation}
      P_0(v) = \frac{v^{\mu}}{\mu+1}, \mbox{ for } v \in [0,1],
  \end{equation}
 where $\mu>-1$ for $P_0(v)$ to be normalizable. We derive it below for completeness and
  explain in detail the subtle approximations involved. Firstly, we re-write $\xi_i(v)$
  as
  \begin{multline}
      \xi_i(v) = \int_0^v dv' (v - v') \Phi_i(v') \\= \int_0^{v^*} dv' (v-v') \frac{P_0(v')}{2} + \int_{v^*}^v dv' (v-v')\Phi_i(v') \\
      = \frac{v^{*^{\mu+1}}}{2(\mu+1)}\left\{ \frac{v  }{\mu+1} - \frac{v^*}{\mu+2} \right\} + {\bar \xi}_i(v), \label{xi1}
  \end{multline}
  where
 \begin{equation}
    {\bar \xi}_i(v) = \int_{v^*}^v dv' (v-v')\Phi_i(v'). \nonumber
 \end{equation}
 and ${\bar \xi}''_i(v)=\Phi_i(v) $. Using Eq.\eqref{xi1} in Eq.\eqref{Probconsecl} we get
 \begin{equation}
    {\bar \xi}''_i(v){\bar \xi}_i(v) + \frac{v^{*^{\mu+1}}}{2(\mu+1)}\left\{ \frac{v  }{\mu+1} - \frac{v^*}{\mu+2} \right\}{\bar \xi_i} \approx P_0(v)v^{*^{\mu+2}}. \label{xi2}
 \end{equation}
Substituting $\bar{\xi}_i \sim v^{*^{\sigma}} v^{\delta}$
 in Eq.{\eqref{xi2}}, with some manipulation we get
 \begin{multline}
     \left( \left(\frac{v}{v^*} \right)^{2\delta - 2} - \left(\frac{v}{v^*} \right)^{\delta - 2} \right) v^{*^{2\sigma + 2 \delta -2}} \\+ \left( \left(\frac{v}{v^*} \right)^{\delta - 1} - \left(\frac{v}{v^*} \right)^{\delta - 2} \right) v^{*^{\sigma + \mu + \delta}}
     \approx \left(\frac{v}{v^*} \right)^{\mu} v^{*^{2\mu + 2}}. \label{xi3}
 \end{multline}
 In the above equation, equating the exponents of $v^*$ on both sides we get the
 relation 
 \begin{equation}
     \sigma + \delta = \mu + 2. \label{sdm}
 \end{equation}
 When $v \gg v^*$ and $\delta > 1$, Eq.\eqref{xi3} may
 be approximated as
 \begin{equation}
    \left(\frac{v}{v^*} \right)^{2\delta - 2} v^{*^{2\sigma + 2 \delta -2}} \approx 
    \left(\frac{v}{v^*} \right)^{\mu} v^{*^{2\mu + 2}},
 \end{equation}
  from which we get $\delta = (\mu + 2)/2$ by equating the exponents of $v/v^*$. Using it in
  Eq.\eqref{sdm},we get $\sigma = (\mu + 2)/2$.
  The above
  relations are valid when $\mu > 0$ as $\delta >1$.
  In a similar way, when $v \gg v^*$ and $\delta < 1$,
  Eq.\eqref{xi3} may be approximated as
  \begin{equation}
      \left(\frac{v}{v^*} \right)^{\delta - 1} v^{*^{\sigma + \mu + \delta}} \approx 
    \left(\frac{v}{v^*} \right)^{\mu} v^{*^{2\mu + 2}}.
  \end{equation}
Equating the coefficients of $v/v^*$ we get $\delta = \mu + 1$ which when substituted in 
Eq.\eqref{sdm} gives $\sigma = 1$. These expressions are valid when $\mu < 0$ as $\delta < 1$.
Summarizing the results,
\begin{eqnarray}
    \Phi_i(v) \sim \left\{ 
    \begin{array}{ll}
    R^{-1/2}v^{(\mu-2)/2} ~~~~~~~~~ \mu \ge 0  \\ 
    R^{-1/(\mu+2)} v^{\mu-1} ~~~~~~~~ \mu < 0  \label{piecl}
\end{array}
    \right. 
\end{eqnarray}
where $R=(R_1 + R_{-1})/4$

 Let us look at the joint distribution on lane $i$, $\Psi_i(v';v)$
 in the limit $v^* \ll 1$. When $v' \ll v^*$ and $v \ll v^*$,
 by the same arguments given for platoon density we note that the
 approximations derived in free-flow limit are valid. 
 Therefore,
 \begin{equation}
    \Psi_i(v';v) \approx \frac{R_i}{4}P_0(v)P_0(v')(v-v').
    \label{P2ecl}
 \end{equation}
 For the case $v'<v^*$ and $v>v^*$, substituting $\Phi_i(v')\approx P_0(v)/2$
 and the expression in Eq.\eqref{piecl} for $\Phi_i(v)$ we get
 \begin{equation}
     \Psi_i(v';v)\approx (R_i^2 + R^2)R^{-\frac{\mu+3}{\mu+2}} v^{\mu}v'^{\mu}
 \end{equation}
 for any $\mu>-1$. For $v' > v^*$ and (obviously) $v > v^*$, we get
\begin{multline}
    \Psi_i(v';v) \\ \sim \left\{ 
    \begin{array}{ll}
    R^{-1/2}v^{\mu}v'^{\mu/2 -1}(v'^{-(\mu+1)} - v^{-(\mu+1)})~~~~(\mu \ge 0) \\ 
    R^{\frac{\mu-1}{\mu+2}}v^{\mu}v'^{\mu -1}(v'^{-1} - v^{-1})~~~~~~~~~~(\mu < 0)  \label{pi2ecl}
\end{array}.
    \right. 
\end{multline}
 The car density  may be obtained by using above derived 
 forms for $\Phi_i(v)$ and $\Psi_i(v';v)$ in Eq.\eqref{carden}.
 For $v<v*$ we get
 \begin{equation}
     G_i(v) \sim  \left( \left(\frac{R_i}{R}\right)^{\frac{\mu+3}{\mu+2}}R_i^{\frac{\mu+1}{\mu+2}} + R^{\frac{\mu+1}{\mu+2}}\right)v^{\mu}
     \label{Gi}
 \end{equation}
for any $\mu>-1$. 
Thus the car density increases enormously from its
initial value for $v < v^*$. Similarly for $v \gg v^*$ and $\mu \ge 0$, we get 
\begin{multline}
G_i(v) \sim R^{-1/2}  v^{\mu/2-1} \\ + R^{-1/2} \left[\frac{v^{-(\mu/2 +2)} - v^{\mu/2 -1}}{(\mu+1)^2} -\frac{v^{\mu/2-1}}{\mu+1}\ln\Big(\frac{1}{v}\Big)\right], \label{gimugt0}
\end{multline}
and for $v \gg v^*$ and $\mu < 0$
\begin{multline}
G_i(v) \sim R^{-1/(\mu+2)}  v^{\mu-1} \\
        + R^{\frac{\mu-1}{\mu+2}} \left[\frac{v^{\mu - 2}(1-v^{\mu+1})}{\mu+1} -\frac{v^{2\mu-2}(1 - v^{\mu})}{\mu}\right]. \label{gimult0}
\end{multline}
Therefore as $v\rightarrow 1$, the car density becomes the same as the
platoon density and it is much smaller than the car density for $v<v^*$.

The average platoon size in lane $i$ may be estimated as
\begin{multline}
    \Lambda_i = \frac{\int_0^{1} dv G_i(v)}{\int_0^{1} dv \Phi_i(v)} 
        = \frac{\int_0^{v^*} dv G_i(v) + \int_{v^*}^1 dv G_i(v)}
        {\int_0^{v^*} dv \Phi_i(v) + \int_{v^*}^{1} dv \Phi_i(v)} \\
        \approx \frac{\int_0^{v^*} dv G_i(v)}{\int_0^{v^*} dv \Phi_i(v)}
    \sim \left( \left(\frac{R_i}{R}\right)^{\frac{\mu+3}{\mu+2}}R_i^{\frac{\mu+1}{\mu+2}} + R^{\frac{\mu+1}{\mu+2}}\right). \label{avgpltsize }
\end{multline}
On RHS of second line of above equation, the contribution to integrals above 
$v > v^*$ has been 
neglected because it is negligible when compared to the contribution from $v < v^*$. 
Thus, the average size of platoons increases significantly and depends on the escape time 
in the lane. If the escape times on both the lanes are of similar magnitude, then
$\Lambda_i \sim  R^{\frac{\mu+1}{\mu+2}}$ for both the lanes.
If $R_j \gg R_i \gg 1$, the leading dependence would be
$\Lambda_i\sim \Lambda_j \sim R_j^{\frac{\mu+1}{\mu+2}}$. 
 Thus, the platoon size in each lane increases as a power-law of the
 largest escape time with the exponent $\frac{\mu+1}{\mu+2}$.
 It may be noted that the form of $\Lambda_i$ is valid for any $\mu > -1$.
 Since the above argument does not use any information about $G_i(v)$, $\Phi_i(v)$ and
 $P_0(v)$ for $v > v^*$ except for the fact that $G_i(v > v^*)$ and $\Phi_i(v > v^*)$ are negligible in comparison to $G_i(v < v^*)$ and $P_i(v < v^*)$ respectively, it
 is valid for any $P_0(v)$ which goes as $v^{\mu}$
 for $v < v^*$. The power-law dependence of $\Lambda_i$ may be related to the
 power-law observed in the single lane case without passing where the platoon
  size increases with time as $t^\frac{\mu+1}{\mu+2}$ from which we may infer in the 
  present case that the platoon size increases until a time scale of escape time is
  reached after which it saturates.
  
The flux in lane $i$ is
\begin{multline}
	J_i = \int_0^1 dv v G_i(v) = \int_0^{v^*} dv v G_i(v) + \int_{v^*}^1 dv v G_i(v) \\ \sim \begin{cases}
		 C_i\frac{{v^*}^{\mu+2}}{\mu+2} + \frac{R^{-1/2}}{\mu/2 + 1}(1-v^{*^{\mu/2+1}}) & \mu \ge 0 \\
		 C_i\frac{{v^*}^{\mu+2}}{\mu+2} + \frac{R^{-1/(\mu+2)}}{\mu + 1}(1-v^{*^{\mu+1}}) & \mu < 0
		\end{cases},
\end{multline}
%
where
\begin{equation}
   C_i = \left(\frac{R_i}{R}\right)^{\frac{\mu+3}{\mu+2}} {R_i}^{\frac{\mu+1}{\mu+2}} + R^{\frac{\mu+1}{\mu+2}}
\end{equation}
and we used the leading terms in \mbox{Eq.\eqref{gimult0}} and \mbox{Eq.\eqref{gimugt0}} to derive the above expression.

\subsection{An example}
Let's plot the flux in each lane when $P_0(v)$ is a beta distribution
 \begin{equation}
	P_{0}(v) = \frac{1}{B(\mu-1,\nu-1)} v^{\mu}(1-v)^{\nu}I_{[0,1]}(v), \label{beta}
\end{equation}
where $I_{[0,1]}(v) = 1$ if $v \in [0,1]$ and 0 otherwise and $B(\mu-1,\nu-1)$ is the beta function with 
parameters $\mu > -1$ and $\nu > -1$. 
For ease of plotting we first write flux
and related quantities in dimensional units. Dimensional quantities are denoted by a bar on the top 
($\bar J$ is dimensional form of $J$).
For $v^* \ge 1$ i.e., $\rho_0 \in \left( 0, \frac{4}{\bar R_1+ \bar R_{-1}}\frac{1}{v_0} \right]$,
 \begin{equation}
  \bar J_i \sim \rho_0 v_0 \frac{J_0}{2} + \frac{\bar R_i-\bar R_{-i}}{8}A_0 (\rho_0 v_0)^2 \label{Jfree}
 \end{equation}
and for $v^* < 1$ i.e., $\rho_0 > \frac{4}{\bar R_1+ \bar R_{-1}}\frac{1}{v_0} $,
\begin{multline}
	\bar J_i \\ \sim \begin{cases} 
		\rho_0 v_0 C_i \frac{v^{*^{\mu+2}}}{\mu+2} + \rho_0 v_0\frac{R^{-1/2}}{\mu/2 + 1}(1-v^{*^{\mu/2+1}}) & \mu \ge 0 \\
		 \rho_0 v_0 C_i\frac{{v^*}^{\mu+2}}{\mu+2} + \rho_0 v_0 \frac{R^{-1/(\mu+2)}}{\mu + 1}(1-v^{*^{\mu+1}}) & \mu < 0
	\end{cases}.
\label{Jcong} 
\end{multline}
%

From \mbox{eq.\eqref{Jfree}} for $\bar J_i$, it may be noted that, when $\bar R_i = \bar R_{-i}$, only the first term remains 
and it is simply the product of arithmetic average of initial density and the expectation value of free-flow speed. 
While \mbox{Eq.\eqref{Jcong}} shows a complicated dependence of $\bar J_i$ on $\rho_0$, it can be shown that $J \sim \rho_0^{\frac{\mu+1}{\mu+2}}$ to leading order in $\rho_0$. Thus the $J_i$ curve flattens as $\rho_0$ becomes greater than $\rho^*$. Below we plot in \mbox{Fig.\ref{fig:Jvsrho}} $\bar J_i$ for $\mu = -0.9$, $\nu = 1$, $\bar R_1 = 0.001 hr$, $\bar R_{-1} = 0.002 hr$. It may be noted that in general kinetic theories are accurate only for low densities. So we depicted only the low density regime in the plot where the transition from free-flow phase in which all the vehicles
experience free-flow to a platoon forming phase in which vehicles with high free-flow speeds form platoons behind slow moving vehicles.

\subsection{Relation to three phase theory}
Kerner et al.\mbox{\cite{Kerner1,Kerner2,Kerner3,Kerner4}}, analyzed a bulk of traffic flow data and noted that the flow has the following pattern across a bottleneck which has been put forward as three-phase theory:
 The traffic flow ahead of 
a bottleneck is in free-flow($F$) phase and at the bottleneck 
the flow undergoes a phase-
transition into a synchronous phase ($S$) in which vehicles face 
congestion but move in a synchronous manner thereby forming
platoons. The $S$-phase spreads downstream as it forms until it
becomes unstable resulting in a congested flow phase($J$)
in which wide moving jams or stop-go waves occur. Thus, one 
observes an $F\rightarrow S$ transition followed by an $S \rightarrow J$ 
transition along the upstream direction across a bottleneck. While
the whole process takes place in a transient state, under suitable
conditions the system may reach a non-equilibrium stationary state
maintaining the same flow pattern across the bottleneck. 

\begin{figure}
    \centering
    \includegraphics[scale=.5]{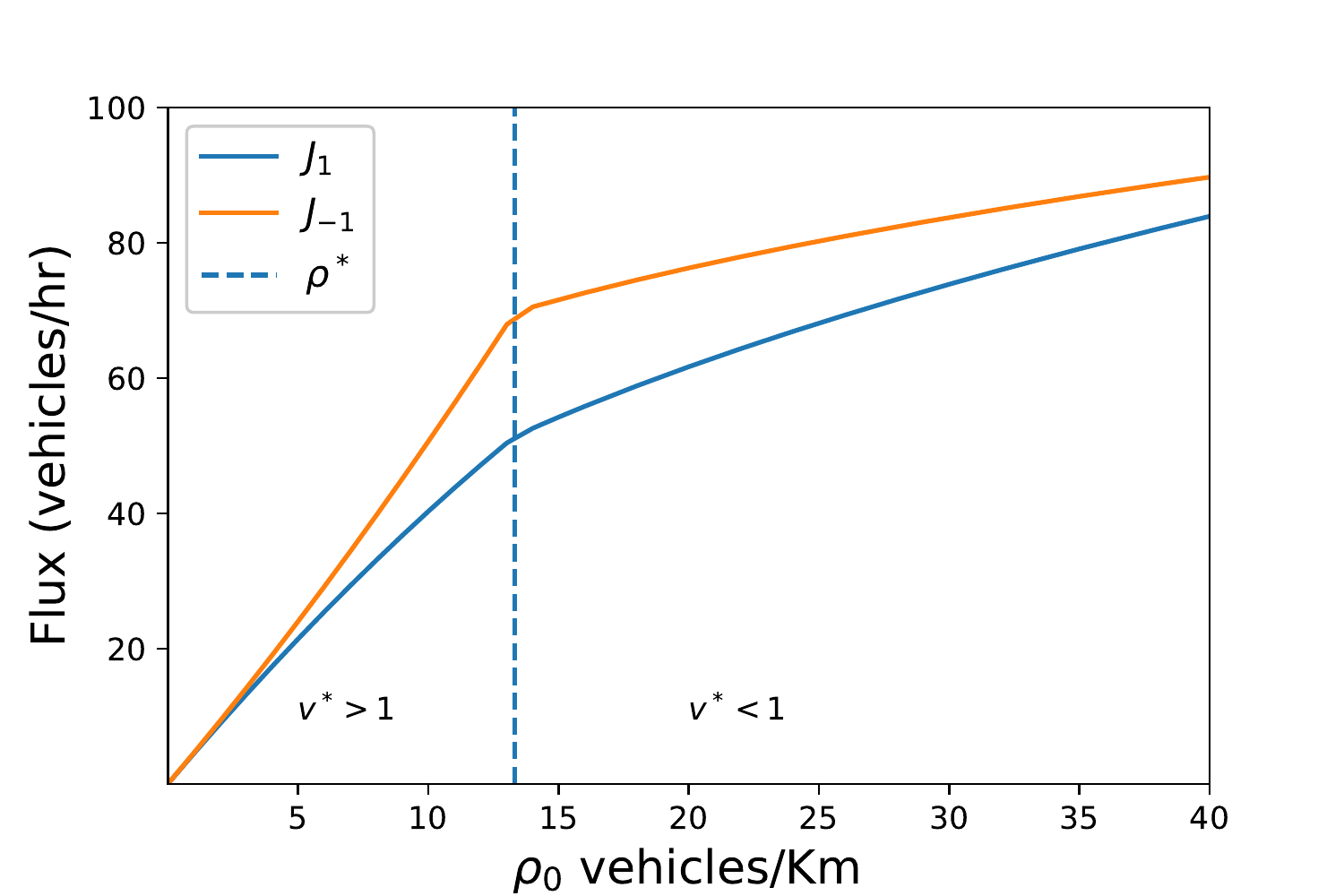}
    \caption{Stationary state fluxes ($J_1, J_{-1}$) on lanes  $1$ and $-1$ as a function of initial global density on the road.For $v^* >= 1$ or $\rho_o \in (0,\rho^*]$, where $\rho^* = \frac{4}{\bar R_1+ \bar R_{-1}}\frac{1}{v_0}$, all the vehicles on both the lanes experience free-flow and the average platoon size is unity. $J_i$ is given by Eq. \eqref{Jfree}. For $v^* < 1$ or $\rho_0 > \rho^*$, platoon size becomes large as given by Eq. \eqref{piecl} and vehicles with high free-flow speeds start feeling congestion. $J_i$ is given by Eq. \eqref{Jcong}. Proportionality constant is computed by constraining $J_i$ to be continuous at $\rho^*$. For the above figure we used $\mu=-0.9$ and $\nu=1$.}
    \label{fig:Jvsrho}
\end{figure}

To relate the present work to three-phase theory, we first have to note a few observations from our 
previous work\mbox{\cite{Ramana2021}} in which we simulated heterogeneous traffic flow on a single lane
using quenched-disordered Newell's car-following model.
 The quenched-disorders are basically the parameters of the model drawn from static probability distributions.
For Newell's model, the parameters are the free-flow speed, the jam density and the backward-wave propagation 
speed. In addition to study of the emergence of power-laws in various quantities like platoon size etc.,
  we also showed that the simulations could
 reproduce the traffic flow pattern as described in the three-phase
  theory. Basically, every slow-moving vehicle forms a 
  moving bottleneck to fast-moving vehicles and thus, ahead a slow-moving vehicle,
  one may find a free-flowing traffic of which 
  it is also a part. Behind the slow-moving vehicle, the fast-moving vehicles form a platoon 
  and thereby experience a $S$-phase which spreads upstream within the platoon
  until the flow becomes unstable resulting in the $J-phase$.
  We described how the formation of $J-phase$ can be explained by the 
  instability created by the reaction time of the drivers. If the 
  drivers reacted instantaneously to any perturbations ahead, the
   $S$-flow would continue up to the end of the
   platoon\mbox{\cite{Ramana2020}}. Further
   it has been observed that even in the $J$-phase, the platoons still exist if the instability is 
   string instability and the power-laws for the 
   platoon size and quantities averaged over the platoon remain the same irrespective
   of whether the system is an $S$-phase or
   $J-phase$. This forms the basis for the validity 
   of the emergent phenomena described by the kinetic theory used in the present work even though the 
   model assumes zero reaction time of drivers to keep the analytical study tractable. In Fig.\ref{fig:Jvsrho}, the flattening of the flux above $\rho^*$ is essentially due to platoon formation
   behind the slow-moving vehicles as described above in relation to three-phase theory.
   
\section{Platoon phase transition} \label{ppt}
The dependence of average platoon size on the escape time could be arrived at 
using just the kinetics of the speed distribution of platoons as shown in
the previous section. However, it is desirable to obtain the size distribution
 of platoons which explains more about the structure of the system.
 Ben-Naim and Krapivsky\cite{Ben1999} derived the stationary size distribution
 for a single lane by further simplifying the kinetic equations for
 cluster size distribution assuming a constant 
 collision rate which they called the Maxwell model.
 Later Ispolatov and Krapivsky\cite{Ispolatov2000} 
 studied a modified version in which only
 next to the leading car is allowed to escape the platoon and
 observed a phase-transition from a phase having smaller platoons of finite size and 
 to a phase having infinitely long platoons (or platoons with size of the order of 
 road length) at $R=1$. They noted that  violation of the 
 sum-rule for vehicle conservation is a signature of the
 phase transition.
Below we formulate a generalized set of equations for kinetics
 of platoon size distribution for the two lane case and study them
 to see if the mentioned violation of the sum rule
 occurs signalling the phase transition. 
  
  Let $P_i(m,t)$ be the density of platoons of size $m$ at time $t$ moving at any
  speed on lane $i$. The kinetic equations followed by 
  $P_i(m,t)$ within the Maxwell model are 
  \begin{multline}
  	\frac{\partial P_i(m,t)}{\partial t} \\= 
  	\begin{cases}
  		\!\begin{aligned}
  			& \frac{1}{R_i} \big( (\alpha(m-1)+1)P_i(m+1,t)  \\
  			& - (\alpha(m-2)+1)P_i(m,t) \big) - P_i(m,t)\rho_i \\
  			& \qquad \qquad \qquad +\frac{1}{2}\sum_l P_i(l,t)P_i(m-l,t) 
  		\end{aligned}           & m \ge 2 \\
  		\!\begin{aligned}
  			& \dfrac{\sum_{n \ge 2}(\alpha(n-2)+1)P_{-i}(n,t)}{R_{-i}} + \dfrac{P_i(2,t)}{R_i} \\
  			& \qquad \qquad \qquad \qquad \qquad \qquad -\rho_i P_i(1,t)
  		\end{aligned}			& m = 1
  	\end{cases}
  \label{pmkt1}
  \end{multline}
In the above equations, $\alpha=1$ implies that each of the follower vehicles in the
 platoon may independently chose to escape at a rate $R_i^{-1}$(as used by Ben-Naim and Krapivsky) and 
 $\alpha=0$ implies only one vehicle, presumably the one next to the leader, may
 escape at a rate $R_i^{-1}$ (as used by Isoplatov and Krapivsky). Thus both the
 cases can be studied in one shot using the above set of equations.
 Further,  $0 < \alpha < 1$ case may be interpreted as the fraction of the follower 
 vehicles that may choose to escape the platoon at a rate $R_i^{-1}$. Eq.\eqref{pmkt1} may be arrived by starting from
 a more general equation for a joint speed-size distribution
 and integrating out the speed (see appendix).
 As stationary 
 state is approached, the equation becomes,
 \begin{multline}
 	P_i(m)\rho_i   = \frac{1}{R_i}  \big( (\alpha(m-1)+1)P_i(m+1)  \\ - (\alpha(m-2)+1)P_i(m)(1-\delta_{m,1}) \big) 
 	+\frac{1}{2}\sum_l P_i(l)P_i(m-l) \\ + \frac{\sum_{m \ge 2}(\alpha(m-2)+1)P_j(m)}{R_j}\delta_{m,1}.  \label{pmkts1}
 \end{multline}
 Further, in the stationary state, the number of vehicles escaping
  lane $1$ per unit time must be equal to the number of vehicles escaping lane $2$ per unit time.
  This can be seen by noting that total number of
  vehicles leaving lane $j$ per unit time is
  \begin{equation}
     \frac{\int_0^1 dv \int_0^v dv' \Psi_i(v';v)}{R_i} =  \frac{\sum_{m \ge 2}(\alpha(m-2)+1)P_i(m)}{R_i} \nonumber
  \end{equation}
  and equating the $LHS$ of Eq.\eqref{Esc1eqcol1} and the $LHS$ of Eq.\eqref{Esc1eqcol2}. Therefore,
 \begin{multline}
     \frac{\sum_{m \ge 2}(\alpha(m-2)+1)P_i(m)}{R_i} \\= \frac{\sum_{m \ge 2}(\alpha(m-2)+1)P_{-i}(m)}{R_{-i}}.
 \end{multline}
 Thus, as the stationary state is approached, the equations in 
 Eq.\eqref{pmkts1} get decoupled and we get
   \begin{multline}
 	P_i(m)\rho_i   = \frac{1}{R_i}   \big( (\alpha(m-1)+1)P_i(m+1) \\ - (\alpha(m-2)+1)P_i(m)(1-\delta_{m,1}) \big) 
 	+\frac{1}{2}\sum_l P_i(l)P_i(m-l) \\ + \frac{\sum_{m \ge 2}(\alpha(m-2)+1)P_i(m)}{R_i}\delta_{m,1}, \label{pmkts2}
 \end{multline}

 which are exactly the equations obtained for a single lane with 
 passing. The $\alpha=0$ and $\alpha=1$ cases have already been studied.
 Readers may refer to Ben-Naim and 
 Krapivsky\cite{Ben1999} and Ispolatov and Krapivsky\cite{Ispolatov2000} for 
 details which we do not repeat here.
 We find some interesting points to note from the generalised
 set of equations which are explained below.
 
\begin{figure}
     \centering
     \includegraphics[scale=0.9]{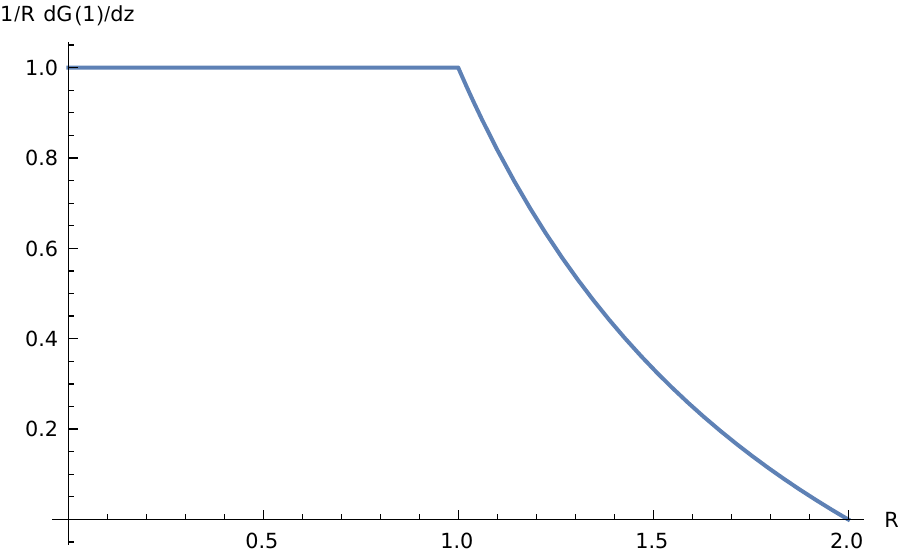}
     \caption{$\frac{d G}{d z}|_{z=1}$ Vs $R$ when $\alpha=0$}
     \label{fig:sumrule2a0}
 \end{figure}

 The above hierarchy of equations are usually solved using a generating 
 function method which is a series of $P_i(m)$ whose coefficients have 
 to be determined using Eq.\eqref{pmkts2}.  Since 
  equations are the same for both the lanes, we do not show the subscript indicating lane for convenience in the below derivation.
Consider the generating function
 \begin{equation}
    \mathcal{G}(z) = \sum_{m=1}^{\infty} (z-1)^m F_m, \label{gf}
\end{equation}
where $F_m = R P(m)$. Using the definition of $\mathcal{G}(z)$ in 
Eq.\eqref{pmkts2} we get
\begin{multline}
\frac{\mathcal{G}^2}{2} + \alpha z (1-z) \frac{d}{dz}\left(\frac{\mathcal{G}}{z} \right) 
+ (1-\alpha)  \frac{(1-z)}{z} \mathcal{G}(z) \\+ (1-2 \alpha) \frac{(1-z)^2}{z}F 
+ \alpha (z-1)R  = 0, \label{gfde1}
\end{multline}
where $F = \sum F_m$. The boundary condition is $\mathcal{G}(1)=0$.
By definition, $\mathcal{G}(0)=-F$ and $\frac{d \mathcal{G}}{d z}|_{z=1} = R$ which are
 sum rules the solution of Eq.\eqref{gfde1} is expected to satisfy. The
 first sum rule is satisfied by definition.
 The second sum rule is related to 
 conservation of vehicles and is satisfied if there
 are no infinite size platoons in the system.When $\alpha=0$, Eq.\eqref{gfde1} 
 becomes a quadratic equation whose physically acceptable solution is
 \begin{equation}
    \mathcal{G}(z;\alpha=0) = \frac{z-1}{z} \left[ 1 - \sqrt{1 - 2 z F} \right]. \label{Ga0}
\end{equation}
  The first sum rule (i.e., $\mathcal{G}(0)=-F$) is trivially followed while 
 the second sum rule (i.e., $\frac{d \mathcal{G}}{d z}|_{z=1} = R$) is followed 
 by $\mathcal{G}(z;\alpha=0)$ for $R \in [0,1]$ (see Fig.\ref{fig:sumrule2a0}).
Violation of the second sum rule above $R=1$ has been 
interpreted as formation 
of an infinitely long platoon (on an infinitely long lane)\cite{Ispolatov2000}.
Thus in this case there is a phase transition from a state
with no platoons or small size platoons to a state with platoons whose size is of
the order of the road length.

\begin{figure}
     \centering
     \includegraphics[scale=0.9]{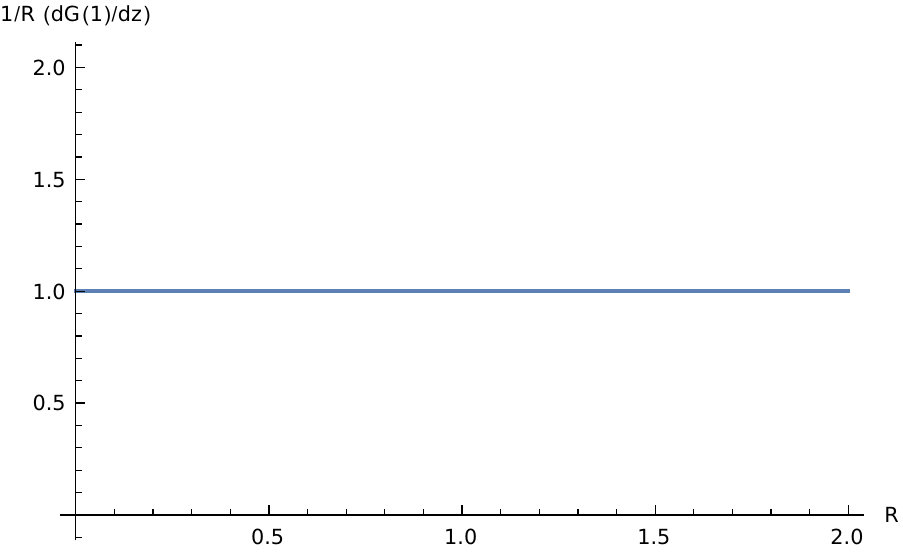}
     \caption{$\frac{d G}{d z}|_{z=1}$ Vs $R$ when $\alpha=1/2$}
     \label{fig:sumrule2a0.5}
 \end{figure}

 Interestingly, Eq.\eqref{pmkts2} has exact solution for $\alpha=1/2$ as well. 
 For $\alpha=1/2$, Eq.\eqref{gfde1} becomes 
\begin{eqnarray}
& &  (1-z) \frac{d \mathcal{G}}{dz} +\mathcal{G}^2 +  (z-1)R  = 0. \label{gfdea0.5}
\end{eqnarray}
Substituting $z=1-x$ and using the transformation $v(x) = \mathcal{G}(x)/x$ and defining
$u(x)$ such that $v(x) = -\frac{1}{u} \frac{d u}{d x}$, we get the Bessel differential
equation for $u(z)$ which is
\begin{equation}
    u''(x) + \frac{u'(x)}{x} -R \frac{u(x)}{x}=0. \label{gfBessel}
\end{equation}
The solution to above equation which satisfies the boundary condition for $\mathcal{G}$ is
\begin{equation}
    u(x) = I_0(2 \sqrt{(R x)}),
\end{equation}
where $I_0$ is the zeroth order Bessel function of first kind. The $\mathcal{G}(z)$ obtained
from above $u(x)$ is 
\begin{equation}
   \mathcal{G}(z) = -\sqrt{R (1-z)} \frac{I_1(2\sqrt{R (1-z)} )}{I_0(2\sqrt{R (1-z)} )} \label{Ga0.5}
\end{equation}
The above solution satisfies both sum rules for all $R$ indicating that there is no phase
transition. See also, fig.\ref{fig:sumrule2a0.5}.
 
 For any $0<\alpha<1$, Eq.\eqref{pmkts2} (a form of Riccati equation) may be re-written as 
 \begin{equation}
     \frac{d^2 u}{d x^2} - \mathcal{R}(x) \frac{d u}{d x} + \mathcal{S}(x) u(x) = 0
      \label{Ric}
 \end{equation}
 where
  \begin{equation}
  \frac{1}{u}\frac{d u}{d x}= \frac{\mathcal{G}(x)}{2 \alpha (1-x)}, \nonumber
  \end{equation}
\begin{equation}
    \mathcal{R}(x) = \frac{(1-2 \alpha)}{\alpha}\frac{1}{1-x} - \frac{1}{x}, \nonumber
\end{equation}
and
\begin{equation}
    \mathcal{S}(x) = \frac{1-2 \alpha}{2 \alpha ^2}\frac{F}{1-x} - \frac{R}{2 \alpha x} \nonumber 
\end{equation}
 and $z=1-x$. Since Eq.\eqref{Ric} is a second order linear equation, one may attempt a 
 power-series solution for $u(x)$. Let
 \begin{equation}
     u(x) = \sum_{\lambda=0}^{\infty} a_{\lambda} x^{k+\lambda}.
 \end{equation}
  To check for the violation of the sum rule, we first note that
 \begin{equation}
  -\left.  \frac{d \mathcal{G}}{d z}\right|_{z\rightarrow 1} = \left.  \frac{d \mathcal{G}}{d x}\right|_{x\rightarrow 0}   = - \frac{2 \alpha a_1}{a_0}. \label{dgx0}
 \end{equation}
 Therefore, it is enough to determine $a_0$ and $a_1$ to obtain the above derivative.
Substituting it in Eq.\eqref{Ric} and comparing coefficients of equal powers of $x$, 
we get the $a_\lambda$ values. The indicial equation gives $k=0$. Equating the coefficient of 
$x^k$ to zero, we get
\begin{equation}
    a_1 = a_0 \frac{R}{2 \alpha} \label{a1}
\end{equation}
 which when used in Eq.\eqref{dgx0}, we get
 \begin{equation}
  \left.  \frac{d \mathcal{G}}{d z}\right|_{z\rightarrow 1} = R. \label{dgxR}
 \end{equation}
 Hence, the sum rule regarding the conservation of vehicles is respected for any $R$ 
 when $\alpha>0$ corroborating with the 
 the result of $\alpha=1/2$. Thus, we see that the phase-transition from
 a phase with small platoons to a phase with large platoons with size of the 
 order of road length occurs only when the number of vehicles leaving the platoon per unit time
 does not increase with the size of the platoon. Another interesting point to note is that if $R_1 > 1 >R_{-1}$ and $\alpha=0$, 
 we see that lane $1$ would have very large size platoons while lane $2$ has small
 platoons in the stationary state.
 
 \section{summary and discussion} \label{sd}
 We formulated a kinetic theory for a dilute two-lane traffic
 within the framework of Ben-Naim Krapivsky model
 and studied the stationary state properties of the system. 
 We find that the platoon velocity distribution and 
 the platoon density on both the lanes get equalized as the stationary 
state is reached whereas the vehicle velocity distribution and the vehicle density 
are different for each lane. Essentially, the lane with larger mean-escape
time would have larger density of vehicles and a larger platoon size.
Similar to the single lane case,
  the vehicles in the system may be characterised into two groups based on a 
  characteristic speed $v^*$ which is inversely proportional to
  mean of escape times on the lanes as $v^* \sim ((R_1+R_{-1})/4)^{-1/(\mu+2)}$ 
  where $\mu$ is the exponent of the quenched disorder in speed $P_0(v)\sim 
  v^{\mu}$.  If $v^* \sim 1$, then all (or a majority of) the vehicles
  in the system enjoy free-flow and the average platoon size is of the order of 
  unity. If $v^* < 1$, then the vehicles with desired speed greater than $v^*$
   experience a congested flow and those with desired speed lesser than $v^*$
    experience free-flow on any lane. In this case, average platoon size is large 
    (proportional to $v^{*^{-(\mu+1)}}$) and is lane dependent. 
     
     We also wrote down the equations for platoon size distribution for the two lane system and showed that the equations can be exactly mapped to the single lane 
     problem. We showed that the phase-transition, from a phase with small platoons 
     to a phase with platoons as large as the road length, happens only if the escape rate of vehicles in a lane is independent of the size of the platoon. 
     
     Overall, the following differences emerge between 
     a homogeneous traffic and a heterogeneous traffic: For a homogeneous traffic on a single 
     lane, i.e., identical vehicles and identical drivers, the vehicle density is 
     homogeneous in the stationary state. When the traffic is heterogeneous, at least
     due to different free-flow speeds adopted by drivers, the density in the 
     stationary state is no more homogeneous. For a single lane road, 
     the stationary state of a heterogeneous traffic is a single platoon with 
     the slowest vehicle as the leader and a large gap ahead of it.
     For a two lane road with lane changing allowed,
     the stationary state of a heterogeneous traffic  has platoons whose 
     average size depends on the harmonic mean of escape-rates between the lanes and the
     exponent $\mu$ of the distribution of the desired speeds of 
     drivers in the small speed limit.

\begin{acknowledgments}
This work was supported by the NYUAD Center for Interacting Urban Networks (CITIES), funded by Tamkeen under the NYUAD Research Institute Award CG001.
\end{acknowledgments}


\appendix

\section{ }
  Let $P_i(m,v,t)$ be the density of platoons of size $(m \ge 2)$ 
  moving with speed $v$ at time $t$ on lane $i$. The kinetic equations followed by it are
  \begin{multline}
      \frac{\partial P_i(m,v,t)}{\partial t} =
      \frac{P_i(m+1,v,t) - P_i(m,v,t)}{R_i}  \\
      - P_i(m,v,t)\int_0^1 dv'|(v-v')| P_i(v',t) \\
      + \sum_l P_i(l,v,t)\int_v^1 dv'(v'-v)P_i(m-l,v',t). \label{pmktv}
  \end{multline}

Using the second fundamental theorem of calculus,
 the collision integrals in the second and
third terms can be re-written as
  \begin{multline}
      \frac{\partial P_i(m,v,t)}{\partial t} =
      \frac{P_i(m+1,v,t) - P_i(m,v,t)}{R_i}  \\
      - \zeta_1(v) P_i(m,v,t)\int_0^1 dv' P_i(v',t)  \\
      + \zeta_2(v)\sum_l P_i(l,v,t)\int_v^1 dv'P_i(m-l,v',t). \label{pmktv2}
  \end{multline}
  Maxwell's model basically assumes that $\zeta_1$ and $\zeta_2$
  are constant. Here we take them to be unity. Thus the 
  equation becomes
   \begin{multline}
      \frac{\partial P_i(m,v,t)}{\partial t} =
      \frac{P_i(m+1,v,t) - P_i(m,v,t)}{R_i} \\ 
      -  P_i(m,v,t)\int_0^1 dv' P_i(v',t) \\
      + \sum_l P_i(l,v,t)\int_v^1 dv' P_i(m-l,v',t). \label{pmktv3}
  \end{multline}
  Integrating over $v$ on both sides and noting that
  \begin{equation}
      \int_0^1 dv P_i(m,v,t) = P_i(m,t)
  \end{equation}
  we get
  \begin{multline}
  \frac{\partial P_i(m,t)}{\partial t} =
      \frac{P_i(m+1,t) - P_i(m,t)}{R_i} 
      - P_i(m,t)\rho_i \\
      +\frac{1}{2}\sum_l P_i(l,t)P_i(m-l,t).
  \end{multline}
The equation followed by $P_i(1,v,t)$ is
\begin{multline}
    \frac{\partial P_i(1,v,t)}{\partial t} = 
    \sum_{m\ge 2} \frac{P_j(m,v,t)}{R_j} + \frac{P_i(2,v,t)}{R_i}  \\
     - P_i(1,v,t)\int_0^1 dv'|(v-v')| P_i(v',t) 
\end{multline}
which upon using Maxwell's approximation and integrating out $v$ 
as described above gives 
\begin{equation}
    \frac{\partial P_i(1,t)}{\partial t}  = 
    \sum_{m\ge 2} \frac{P_j(m,t)}{R_j} + \frac{P_i(2,t)}{R_i} 
     - P_i(1,t)\rho_i 
\end{equation}

\section*{Bibliography}
\nocite{*}
\bibliography{aipsamp}

\providecommand{\noopsort}[1]{}\providecommand{\singleletter}[1]{#1}%
\begin{thebibliography}{27}%
\makeatletter
\providecommand \@ifxundefined [1]{%
 \@ifx{#1\undefined}
}%
\providecommand \@ifnum [1]{%
 \ifnum #1\expandafter \@firstoftwo
 \else \expandafter \@secondoftwo
 \fi
}%
\providecommand \@ifx [1]{%
 \ifx #1\expandafter \@firstoftwo
 \else \expandafter \@secondoftwo
 \fi
}%
\providecommand \natexlab [1]{#1}%
\providecommand \enquote  [1]{``#1''}%
\providecommand \bibnamefont  [1]{#1}%
\providecommand \bibfnamefont [1]{#1}%
\providecommand \citenamefont [1]{#1}%
\providecommand \href@noop [0]{\@secondoftwo}%
\providecommand \href [0]{\begingroup \@sanitize@url \@href}%
\providecommand \@href[1]{\@@startlink{#1}\@@href}%
\providecommand \@@href[1]{\endgroup#1\@@endlink}%
\providecommand \@sanitize@url [0]{\catcode `\\12\catcode `\$12\catcode
  `\&12\catcode `\#12\catcode `\^12\catcode `\_12\catcode `\%12\relax}%
\providecommand \@@startlink[1]{}%
\providecommand \@@endlink[0]{}%
\providecommand \url  [0]{\begingroup\@sanitize@url \@url }%
\providecommand \@url [1]{\endgroup\@href {#1}{\urlprefix }}%
\providecommand \urlprefix  [0]{URL }%
\providecommand \Eprint [0]{\href }%
\providecommand \doibase [0]{http://dx.doi.org/}%
\providecommand \selectlanguage [0]{\@gobble}%
\providecommand \bibinfo  [0]{\@secondoftwo}%
\providecommand \bibfield  [0]{\@secondoftwo}%
\providecommand \translation [1]{[#1]}%
\providecommand \BibitemOpen [0]{}%
\providecommand \bibitemStop [0]{}%
\providecommand \bibitemNoStop [0]{.\EOS\space}%
\providecommand \EOS [0]{\spacefactor3000\relax}%
\providecommand \BibitemShut  [1]{\csname bibitem#1\endcsname}%
\let\auto@bib@innerbib\@empty
\bibitem [{\citenamefont {Habibovic}\ and\ \citenamefont {Chen}(2021)}]{Habib}%
  \BibitemOpen
  \bibfield  {author} {\bibinfo {author} {\bibfnamefont {A.}~\bibnamefont
  {Habibovic}}\ and\ \bibinfo {author} {\bibfnamefont {L.}~\bibnamefont
  {Chen}},\ }\href@noop {} {\emph {\bibinfo {title} {Connected Automated
  Vehicles: Technologies, Developments, and Trends}}}\ (\bibinfo  {publisher}
  {Elsevier},\ \bibinfo {year} {2021})\BibitemShut {NoStop}%
\bibitem [{\citenamefont {Stevens}\ and\ \citenamefont
  {Hopkin}(2012)}]{Stevens}%
  \BibitemOpen
  \bibfield  {author} {\bibinfo {author} {\bibfnamefont {A.}~\bibnamefont
  {Stevens}}\ and\ \bibinfo {author} {\bibfnamefont {J.}~\bibnamefont
  {Hopkin}},\ }\bibfield  {title} {\enquote {\bibinfo {title} {Benefits and
  deployment opportunities for vehicle/roadside cooperative its},}\ }\href@noop
  {} {\bibfield  {journal} {\bibinfo  {journal} {IET and ITS Conference on Road
  Transport Information and Control (RTIC 2012)}\ ,\ \bibinfo {pages} {1--6}}
  (\bibinfo {year} {2012})}\BibitemShut {NoStop}%
\bibitem [{SAE()}]{SAE_J3216}%
  \BibitemOpen
  \bibfield  {title} {\enquote {\bibinfo {title} {Taxonomy and definitions for
  terms related to driving automation systems for on-road motor vehicles},}\
  }\href@noop {} {\bibinfo  {journal} {On-Road Automated Driving (ORAD)
  committee, SAE J3216 Standard}\ }\BibitemShut {NoStop}%
\bibitem [{\citenamefont {Xu}\ \emph {et~al.}(2021)\citenamefont {Xu},
  \citenamefont {Guo}, \citenamefont {Han}, \citenamefont {Xia}, \citenamefont
  {Xiang},\ and\ \citenamefont {Ma}}]{Opencda}%
  \BibitemOpen
\bibfield  {journal} {  }\bibfield  {author} {\bibinfo {author} {\bibfnamefont
  {R.}~\bibnamefont {Xu}}, \bibinfo {author} {\bibfnamefont {Y.}~\bibnamefont
  {Guo}}, \bibinfo {author} {\bibfnamefont {X.}~\bibnamefont {Han}}, \bibinfo
  {author} {\bibfnamefont {X.}~\bibnamefont {Xia}}, \bibinfo {author}
  {\bibfnamefont {H.}~\bibnamefont {Xiang}}, \ and\ \bibinfo {author}
  {\bibfnamefont {J.}~\bibnamefont {Ma}},\ }\bibfield  {title} {\enquote
  {\bibinfo {title} {Opencda: An open cooperative driving automation
  frameworkintegrated with co-simulation},}\ }\href@noop {} {\bibfield
  {journal} {\bibinfo  {journal} {arXiv preprint arXiv:2107.06260}\ } (\bibinfo
  {year} {2021})}\BibitemShut {NoStop}%
\bibitem [{\citenamefont {Krug}\ and\ \citenamefont
  {Ferrari}(1996)}]{Krug1996}%
  \BibitemOpen
  \bibfield  {author} {\bibinfo {author} {\bibfnamefont {J.}~\bibnamefont
  {Krug}}\ and\ \bibinfo {author} {\bibfnamefont {P.~A.}\ \bibnamefont
  {Ferrari}},\ }\bibfield  {title} {\enquote {\bibinfo {title} {Phase
  transitions in driven diffusive systems with random rates},}\ }\href@noop {}
  {\bibfield  {journal} {\bibinfo  {journal} {Journal of Physics A:
  Mathematical and General}\ }\textbf {\bibinfo {volume} {29}},\ \bibinfo
  {pages} {L465} (\bibinfo {year} {1996})}\BibitemShut {NoStop}%
\bibitem [{\citenamefont {Evans}(1996)}]{Evans1996}%
  \BibitemOpen
  \bibfield  {author} {\bibinfo {author} {\bibfnamefont {M.}~\bibnamefont
  {Evans}},\ }\bibfield  {title} {\enquote {\bibinfo {title} {Bose-einstein
  condensation in disordered exclusion models and relation to traffic flow},}\
  }\href@noop {} {\bibfield  {journal} {\bibinfo  {journal} {EPL (Europhysics
  Letters)}\ }\textbf {\bibinfo {volume} {36}},\ \bibinfo {pages} {13}
  (\bibinfo {year} {1996})}\BibitemShut {NoStop}%
\bibitem [{\citenamefont {Ktitarev}, \citenamefont {Chowdhury},\ and\
  \citenamefont {Wolf}(1997)}]{Ktitarev1997}%
  \BibitemOpen
  \bibfield  {author} {\bibinfo {author} {\bibfnamefont {D.~V.}\ \bibnamefont
  {Ktitarev}}, \bibinfo {author} {\bibfnamefont {D.}~\bibnamefont {Chowdhury}},
  \ and\ \bibinfo {author} {\bibfnamefont {D.~E.}\ \bibnamefont {Wolf}},\
  }\bibfield  {title} {\enquote {\bibinfo {title} {Stochastic traffic model
  with random deceleration probabilities: queueing and power-law gap
  distribution},}\ }\href@noop {} {\bibfield  {journal} {\bibinfo  {journal}
  {Journal of Physics A: Mathematical and General}\ }\textbf {\bibinfo {volume}
  {30}},\ \bibinfo {pages} {L221} (\bibinfo {year} {1997})}\BibitemShut
  {NoStop}%
\bibitem [{\citenamefont {Bengrine}\ \emph {et~al.}(1999)\citenamefont
  {Bengrine}, \citenamefont {Benyoussef}, \citenamefont {Ez-Zahraouy},
  \citenamefont {Krug}, \citenamefont {Loulidi},\ and\ \citenamefont
  {Mhirech}}]{Bengrine1999}%
  \BibitemOpen
  \bibfield  {author} {\bibinfo {author} {\bibfnamefont {M.}~\bibnamefont
  {Bengrine}}, \bibinfo {author} {\bibfnamefont {A.}~\bibnamefont
  {Benyoussef}}, \bibinfo {author} {\bibfnamefont {H.}~\bibnamefont
  {Ez-Zahraouy}}, \bibinfo {author} {\bibfnamefont {J.}~\bibnamefont {Krug}},
  \bibinfo {author} {\bibfnamefont {M.}~\bibnamefont {Loulidi}}, \ and\
  \bibinfo {author} {\bibfnamefont {F.}~\bibnamefont {Mhirech}},\ }\bibfield
  {title} {\enquote {\bibinfo {title} {A simulation study of an asymmetric
  exclusion model with open boundaries and random rates},}\ }\href@noop {}
  {\bibfield  {journal} {\bibinfo  {journal} {Journal of Physics A:
  Mathematical and General}\ }\textbf {\bibinfo {volume} {32}},\ \bibinfo
  {pages} {2527} (\bibinfo {year} {1999})}\BibitemShut {NoStop}%
\bibitem [{\citenamefont {Chowdhury}, \citenamefont {Santen},\ and\
  \citenamefont {Schadschneider}(2000)}]{DC2000}%
  \BibitemOpen
  \bibfield  {author} {\bibinfo {author} {\bibfnamefont {D.}~\bibnamefont
  {Chowdhury}}, \bibinfo {author} {\bibfnamefont {L.}~\bibnamefont {Santen}}, \
  and\ \bibinfo {author} {\bibfnamefont {A.}~\bibnamefont {Schadschneider}},\
  }\bibfield  {title} {\enquote {\bibinfo {title} {Statistical physics of
  vehicular traffic and some related systems},}\ }\href@noop {} {\bibfield
  {journal} {\bibinfo  {journal} {Physics Reports}\ }\textbf {\bibinfo {volume}
  {329}},\ \bibinfo {pages} {199--329} (\bibinfo {year} {2000})}\BibitemShut
  {NoStop}%
\bibitem [{\citenamefont {Barma}(2006)}]{Barma2006}%
  \BibitemOpen
  \bibfield  {author} {\bibinfo {author} {\bibfnamefont {M.}~\bibnamefont
  {Barma}},\ }\bibfield  {title} {\enquote {\bibinfo {title} {Driven diffusive
  systems with disorder},}\ }\href@noop {} {\bibfield  {journal} {\bibinfo
  {journal} {Physica A: Statistical Mechanics and its Applications}\ }\textbf
  {\bibinfo {volume} {372}},\ \bibinfo {pages} {22--33} (\bibinfo {year}
  {2006})}\BibitemShut {NoStop}%
\bibitem [{\citenamefont {Ramana}\ and\ \citenamefont
  {Jabari}(2020)}]{Ramana2020}%
  \BibitemOpen
  \bibfield  {author} {\bibinfo {author} {\bibfnamefont {A.~S.~V.}\
  \bibnamefont {Ramana}}\ and\ \bibinfo {author} {\bibfnamefont {S.~E.}\
  \bibnamefont {Jabari}},\ }\bibfield  {title} {\enquote {\bibinfo {title}
  {Traffic flow with multiple quenched disorders},}\ }\href@noop {} {\bibfield
  {journal} {\bibinfo  {journal} {Physical Review E}\ }\textbf {\bibinfo
  {volume} {101}},\ \bibinfo {pages} {052127} (\bibinfo {year}
  {2020})}\BibitemShut {NoStop}%
\bibitem [{\citenamefont {Ramana}\ and\ \citenamefont
  {Jabari}(2021)}]{Ramana2021}%
  \BibitemOpen
  \bibfield  {author} {\bibinfo {author} {\bibfnamefont {A.~S.~V.}\
  \bibnamefont {Ramana}}\ and\ \bibinfo {author} {\bibfnamefont {S.~E.}\
  \bibnamefont {Jabari}},\ }\bibfield  {title} {\enquote {\bibinfo {title}
  {Power laws and phase transitions in heterogenous car following with reaction
  times},}\ }\href@noop {} {\bibfield  {journal} {\bibinfo  {journal} {Physical
  Review E}\ }\textbf {\bibinfo {volume} {103}},\ \bibinfo {pages} {032202}
  (\bibinfo {year} {2021})}\BibitemShut {NoStop}%
\bibitem [{\citenamefont {Ben-Naim}, \citenamefont {Krapivsky},\ and\
  \citenamefont {Redner}(1994)}]{Ben1994}%
  \BibitemOpen
  \bibfield  {author} {\bibinfo {author} {\bibfnamefont {E.}~\bibnamefont
  {Ben-Naim}}, \bibinfo {author} {\bibfnamefont {P.~L.}\ \bibnamefont
  {Krapivsky}}, \ and\ \bibinfo {author} {\bibfnamefont {S.}~\bibnamefont
  {Redner}},\ }\bibfield  {title} {\enquote {\bibinfo {title} {Kinetics of
  clustering in traffic flows},}\ }\href@noop {} {\bibfield  {journal}
  {\bibinfo  {journal} {Physical Review E}\ }\textbf {\bibinfo {volume} {50}},\
  \bibinfo {pages} {822} (\bibinfo {year} {1994})}\BibitemShut {NoStop}%
\bibitem [{\citenamefont {Jabari}, \citenamefont {Zheng},\ and\ \citenamefont
  {Liu}(2014)}]{Saif2014}%
  \BibitemOpen
  \bibfield  {author} {\bibinfo {author} {\bibfnamefont {S.~E.}\ \bibnamefont
  {Jabari}}, \bibinfo {author} {\bibfnamefont {J.}~\bibnamefont {Zheng}}, \
  and\ \bibinfo {author} {\bibfnamefont {H.~X.}\ \bibnamefont {Liu}},\
  }\bibfield  {title} {\enquote {\bibinfo {title} {A probabilistic stationary
  speed--density relation based on newell’s simplified car-following
  model},}\ }\href@noop {} {\bibfield  {journal} {\bibinfo  {journal}
  {Transportation Research Part B: Methodological}\ }\textbf {\bibinfo {volume}
  {68}},\ \bibinfo {pages} {205--223} (\bibinfo {year} {2014})}\BibitemShut
  {NoStop}%
\bibitem [{\citenamefont {Jabari}\ \emph {et~al.}(2018)\citenamefont {Jabari},
  \citenamefont {Zheng}, \citenamefont {Liu},\ and\ \citenamefont
  {Filipovska}}]{Saif2018}%
  \BibitemOpen
  \bibfield  {author} {\bibinfo {author} {\bibfnamefont {S.}~\bibnamefont
  {Jabari}}, \bibinfo {author} {\bibfnamefont {F.}~\bibnamefont {Zheng}},
  \bibinfo {author} {\bibfnamefont {H.}~\bibnamefont {Liu}}, \ and\ \bibinfo
  {author} {\bibfnamefont {M.}~\bibnamefont {Filipovska}},\ }\bibfield  {title}
  {\enquote {\bibinfo {title} {Stochastic {L}agrangian modeling of traffic
  dynamics},}\ }in\ \href@noop {} {\emph {\bibinfo {booktitle} {The 97th Annual
  Meeting of the Transportation Research Board, Washington D.C}}}\ (\bibinfo
  {year} {2018})\ pp.\ \bibinfo {pages} {18--04170}\BibitemShut {NoStop}%
\bibitem [{\citenamefont {Helbing}(2001)}]{Helbing2001}%
  \BibitemOpen
  \bibfield  {author} {\bibinfo {author} {\bibfnamefont {D.}~\bibnamefont
  {Helbing}},\ }\bibfield  {title} {\enquote {\bibinfo {title} {Traffic and
  related self-driven many-particle systems},}\ }\href@noop {} {\bibfield
  {journal} {\bibinfo  {journal} {Reviews of modern physics}\ }\textbf
  {\bibinfo {volume} {73}},\ \bibinfo {pages} {1067} (\bibinfo {year}
  {2001})}\BibitemShut {NoStop}%
\bibitem [{\citenamefont {Han}\ \emph {et~al.}(2021)\citenamefont {Han},
  \citenamefont {Shi}, \citenamefont {Chen}, \citenamefont {Li},\ and\
  \citenamefont {Wang}}]{Han2021}%
  \BibitemOpen
  \bibfield  {author} {\bibinfo {author} {\bibfnamefont {J.}~\bibnamefont
  {Han}}, \bibinfo {author} {\bibfnamefont {H.}~\bibnamefont {Shi}}, \bibinfo
  {author} {\bibfnamefont {L.}~\bibnamefont {Chen}}, \bibinfo {author}
  {\bibfnamefont {H.}~\bibnamefont {Li}}, \ and\ \bibinfo {author}
  {\bibfnamefont {X.}~\bibnamefont {Wang}},\ }\bibfield  {title} {\enquote
  {\bibinfo {title} {The car-following model and its applications in the v2x
  environment: A historical review},}\ }\href@noop {} {\bibfield  {journal}
  {\bibinfo  {journal} {Future Internet}\ }\textbf {\bibinfo {volume} {14}},\
  \bibinfo {pages} {14} (\bibinfo {year} {2021})}\BibitemShut {NoStop}%
\bibitem [{\citenamefont {Tanimoto}(2019)}]{Tanimoto1}%
  \BibitemOpen
  \bibfield  {author} {\bibinfo {author} {\bibfnamefont {J.}~\bibnamefont
  {Tanimoto}},\ }\href@noop {} {\emph {\bibinfo {title} {Evolutionary Games
  with Sociophysics: Analysis of Traffic Flow and Epidemics}}}\ (\bibinfo
  {publisher} {Springer.},\ \bibinfo {year} {2019})\BibitemShut {NoStop}%
\bibitem [{\citenamefont {Tanimoto}(2015)}]{Tanimoto2}%
  \BibitemOpen
  \bibfield  {author} {\bibinfo {author} {\bibfnamefont {J.}~\bibnamefont
  {Tanimoto}},\ }\href@noop {} {\emph {\bibinfo {title} {Fundamentals of
  Evolutionary Game Theory and its Applications}}}\ (\bibinfo  {publisher}
  {Springer.},\ \bibinfo {year} {2015})\BibitemShut {NoStop}%
\bibitem [{\citenamefont {Ben-Naim}\ and\ \citenamefont
  {Krapivsky}(1997)}]{Ben1997}%
  \BibitemOpen
  \bibfield  {author} {\bibinfo {author} {\bibfnamefont {E.}~\bibnamefont
  {Ben-Naim}}\ and\ \bibinfo {author} {\bibfnamefont {P.}~\bibnamefont
  {Krapivsky}},\ }\bibfield  {title} {\enquote {\bibinfo {title} {Stationary
  velocity distributions in traffic flows},}\ }\href@noop {} {\bibfield
  {journal} {\bibinfo  {journal} {Physical Review E}\ }\textbf {\bibinfo
  {volume} {56}},\ \bibinfo {pages} {6680} (\bibinfo {year}
  {1997})}\BibitemShut {NoStop}%
\bibitem [{\citenamefont {Ben-Naim}\ and\ \citenamefont
  {Krapivsky}(1998)}]{Ben1998}%
  \BibitemOpen
  \bibfield  {author} {\bibinfo {author} {\bibfnamefont {E.}~\bibnamefont
  {Ben-Naim}}\ and\ \bibinfo {author} {\bibfnamefont {P.}~\bibnamefont
  {Krapivsky}},\ }\bibfield  {title} {\enquote {\bibinfo {title} {Steady-state
  properties of traffic flows},}\ }\href@noop {} {\bibfield  {journal}
  {\bibinfo  {journal} {Journal of Physics A: Mathematical and General}\
  }\textbf {\bibinfo {volume} {31}},\ \bibinfo {pages} {8073} (\bibinfo {year}
  {1998})}\BibitemShut {NoStop}%
\bibitem [{\citenamefont {Ben-Naim}\ and\ \citenamefont
  {Krapivsky}(1999)}]{Ben1999}%
  \BibitemOpen
  \bibfield  {author} {\bibinfo {author} {\bibfnamefont {E.}~\bibnamefont
  {Ben-Naim}}\ and\ \bibinfo {author} {\bibfnamefont {P.}~\bibnamefont
  {Krapivsky}},\ }\bibfield  {title} {\enquote {\bibinfo {title} {Maxwell model
  of traffic flows},}\ }\href@noop {} {\bibfield  {journal} {\bibinfo
  {journal} {Physical Review E}\ }\textbf {\bibinfo {volume} {59}},\ \bibinfo
  {pages} {88} (\bibinfo {year} {1999})}\BibitemShut {NoStop}%
\bibitem [{\citenamefont {Ispolatov}\ and\ \citenamefont
  {Krapivsky}(2000)}]{Ispolatov2000}%
  \BibitemOpen
  \bibfield  {author} {\bibinfo {author} {\bibfnamefont {I.}~\bibnamefont
  {Ispolatov}}\ and\ \bibinfo {author} {\bibfnamefont {P.}~\bibnamefont
  {Krapivsky}},\ }\bibfield  {title} {\enquote {\bibinfo {title} {Phase
  transition in a traffic model with passing},}\ }\href@noop {} {\bibfield
  {journal} {\bibinfo  {journal} {Physical Review E}\ }\textbf {\bibinfo
  {volume} {62}},\ \bibinfo {pages} {5935} (\bibinfo {year}
  {2000})}\BibitemShut {NoStop}%
\bibitem [{\citenamefont {Kerner}(2004)}]{Kerner1}%
  \BibitemOpen
  \bibfield  {author} {\bibinfo {author} {\bibfnamefont {B.~S.}\ \bibnamefont
  {Kerner}},\ }\href@noop {} {\emph {\bibinfo {title} {The physics of
  traffic}}}\ (\bibinfo  {publisher} {Springer},\ \bibinfo {year}
  {2004})\BibitemShut {NoStop}%
\bibitem [{\citenamefont {Kerner}(2017)}]{Kerner2}%
  \BibitemOpen
  \bibfield  {author} {\bibinfo {author} {\bibfnamefont {B.~S.}\ \bibnamefont
  {Kerner}},\ }\href@noop {} {\emph {\bibinfo {title} {Breakdown in traffic
  networks}}}\ (\bibinfo  {publisher} {Springer},\ \bibinfo {year}
  {2017})\BibitemShut {NoStop}%
\bibitem [{\citenamefont {Kerner}(2021)}]{Kerner3}%
  \BibitemOpen
  \bibfield  {author} {\bibinfo {author} {\bibfnamefont {B.~S.}\ \bibnamefont
  {Kerner}},\ }\href@noop {} {\emph {\bibinfo {title} {Understanding Real
  Traffic: Paradigm Shift in Transportation Science}}}\ (\bibinfo  {publisher}
  {Springer, Cham, Switzerland},\ \bibinfo {year} {2021})\BibitemShut {NoStop}%
\bibitem [{\citenamefont {H.~Rehborn}(2021)}]{Kerner4}%
  \BibitemOpen
  \bibfield  {author} {\bibinfo {author} {\bibfnamefont {S.~K.}\ \bibnamefont
  {H.~Rehborn}, \bibfnamefont {M.~Koller}},\ }\href@noop {} {\emph {\bibinfo
  {title} {Data-Driven Traffic Engineering}}}\ (\bibinfo  {publisher}
  {Elsevier, Amsterdam.},\ \bibinfo {year} {2021})\BibitemShut {NoStop}%
\end{thebibliography}%



\end{document}